\documentclass{article}
\usepackage{graphicx}
\usepackage{epsfig}
\usepackage{amsfonts}
\usepackage{slashed}

 \usepackage{amssymb}
\usepackage{amsmath}
\usepackage{color}
\usepackage{hyperref}
\usepackage{graphicx}       
\usepackage{dcolumn}        
\usepackage{bm}             
\usepackage{amssymb}
\usepackage{amstext}
\usepackage{tensor}
\usepackage{cite}

 \usepackage{amssymb}

\date{\today}

\newcommand{\insertplot}[5]{\begin{figure}
 \hfill\hbox to 0.05in{\vbox to #5in{\vfill
 \inputplot{#1}{#4}{#5}}\hfill}
 \hfill\vspace{-.1in}
 \caption{#2}\label{#3}
 \end{figure}}
 \newcommand{\inputplot}[3]{
 \special{ps: plotfile #1}
}
\newcounter{fig}

\newcommand{\beq}{\begin{equation}}
\newcommand{\eeq}{\end{equation}}

\begin{document}
\title{\Large
{\bf
Spinning gauged boson and Dirac stars:\\ a comparative study}
}
 \vspace{1.5truecm}

\author{
{\large }
{\  C. Herdeiro}$^{1}$,
{\ I. Perapechka}$^{2}$,
{  E. Radu}$^{1}$,
and
{ Ya. Shnir}$^{3}$
\\
\\
$^{1}${\small  Departamento de Matemática da Universidade de Aveiro}
\\
 {\small
Center for Research and Development in Mathematics and Applications -- CIDMA}
\\
 {\small
Campus de Santiago, 3810-183 Aveiro, Portugal
}
\\
$^{2}${\small Department of Theoretical Physics and Astrophysics, Belarusian State
University,
}
\\
{\small
Nezavisimosti Avenue 4, Minsk 220004, Belarus}
\\
$^{3}${\small
BLTP, JINR,
Joliot-Curie 6, Dubna 141980, Moscow Region, Russia}
}

\date{November 2021}

\maketitle

\begin{abstract}
Scalar boson stars and Dirac stars are solitonic solutions of the Einstein--Klein-Gordon and Einstein-Dirac classical equations, respectively. Despite the different bosonic $vs.$ fermionic nature of the matter field, these solutions to the classical field equations have been shown to have qualitatively similar features~\cite{Herdeiro:2017fhv}. In particular, for spinning stars the most fundamental configurations can be in both cases toroidal, unlike spinning Proca stars that are spheroidal~\cite{Herdeiro:2019mbz}. In this paper we gauge the scalar and Dirac fields, by minimally coupling them to standard electromagnetism. We explore the impact of the gauge coupling on the resulting solutions.  One of the most  relevant difference concerns the gyromagnetic ratio, which for the scalar stars takes values around 1, whereas for Dirac stars takes values around 2.

\end{abstract}


\section{Introduction and motivation}

In a recent paper \cite{Herdeiro:2019mbz}, we have performed
a comparative analysis of the spinning solitons arising in
the Einstein-Klein-Gordon, Einstein-Dirac and Einstein-Proca models.
Using numerical methods,
particle-like solutions were found and their basic properties analysed. This analysis has indicated a, \textit{a priori} non-obvious,
  high degree of universality between the different models, and in particular with some features holding for
    both bosonic and Dirac stars. 
		Amongst these, we have noticed that both scalar and Dirac stars can present a distinctive toroidal morphology, contrasting with the spheroidal morphology of Proca stars~\cite{Herdeiro:2019mbz}.

Bosonic stars (scalar or vector) can in principle be macroscopic objects, corresponding to many bosons in the same quantum state, to justify the classical description. In fact these models have been widely considered in astrophysical contexts, for instance as black hole mimickers - see $e.g.$~\cite{Bustillo:2020syj,Herdeiro:2021lwl} for some recent discussions. The status of Dirac stars is less clear. Still, one may entertain the possibility that both models could be an approximate description for microscopic objects, eventually relevant in the early Universe. In such context, a differentiated phenomenology could arise, 
$e.g.$ from their interaction with electromagnetic fields.

With this physical motivation, besides the intrinsic interest in understanding self-gravitating solitons in simple physical models, in this paper we  extend
the results in~\cite{Herdeiro:2017fhv} to the case of gauged matter fields. We shall focus on the scalar and Dirac cases, and consider the Einstein--Klein-Gordon and Einstein--Dirac equations minimally
coupled to an electromagnetic field.
 We shall construct the electrically charged, spinning particle-like solutions,
which generalize the neutral solitons in  \cite{Herdeiro:2019mbz},
 and describe some of their properties, in particular their gyromagnetic ratio.

Static, spherically
symmetric \textit{neutral} scalar ($s=0$) boson stars have been known for more than half a century~\cite{Kaup:1968zz,Ruffini:1969qy}. Their gauged generalization, on the other hand,   were only discussed for the first time
in~\cite{Jetzer:1989av}, and more recently in~\cite{Pugliese:2013gsa}.
Moreover, \textit{spinning} boson stars were first constructed in~\cite{Schunck:1996wa,Yoshida:1997qf} for free and in~\cite{Kleihaus:2005me,Kleihaus:2007vk} for  self-interacting (with a $Q$-ball type potential) complex scalar fields. Finally, \textit{spinning gauged} boson stars have also been constructed for free~\cite{Delgado:2016jxq} and  self-interacting (with a $Q$-ball type potential) complex scalar fields~\cite{Brihaye:2009dx,Collodel:2019ohy}.\footnote{One should mention, however,
the existence of a large literature on spinning
gauged
solitons in models with scalar field multiplets and
 non-Abelian gauge fields,
see $e.g.$ \cite{Kleihaus:2016rgf} and references therein.
}

Spherical, neutral Dirac ($s=1/2$) stars were first constructed in~\cite{Finster:1998ws} and their gauged version in~\cite{Finster:1998ux}.
The latter are solutions of the Einstein-Maxwell-Dirac equations with \textit{two} gauged fermions, with opposite spins, in order to satisfy spherical symmetry. Thus, both the neutral and the charged spherical solutions require at least two Dirac fields. Solutions with a \textit{single} Dirac field were constructed as spinning (neutral) solutions in  \cite{Herdeiro:2019mbz}  for the first time, in a model that, therefore, does not admit static stars. So far, no construction of spinning gauged Dirac stars has appeared in the literature.

The solutions reported below - the
gauged generalizations of the spinning solitons with
$s=0,1/2$
in~\cite{Herdeiro:2019mbz} -
possess a  nonzero ADM mass, electric charge and a magnetic dipole moment, similarly to the well known Kerr-Newman black hole of electrovacuum. The latter has a gyromagnetic ratio of $g=2$~\cite{Carter:1968rr} so one may inquire about the gyromagnetic ratio of these solutions, which is an important quantity in particle physics.
Indeed, the experimental value of the quantity $g-2$, which is known with an incredible accuracy,  is a very precise
test of the Standard Model and possible deviations from the theoretical value may be a smoking gun for new physics.
In fact, the recently discussed possible disagreement between theory and experiment for $(g-2)_{\rm muon}$ 
is one of the most hotly debated topics in the particle physics community~\cite{Morel2021}.
It has also been suggested that strong gravitational interaction may also play an important role in
very precise calculations of the corrections to the gyromagnetic ratio
\cite{Garfinkle:1990ib,Khriplovich:1997ni,Pfister:2002dz,Pfister:2002}. 
Concerning the solitons discussed herein, as we shall see there is a clear quantitative difference between $g$ for scalar and Dirac gauged spinning stars.

This paper is organized as follows. In Section~\ref{sec2} we exhibit the models, discuss their equations of motion, relevant physical quantities and the Ansatze that will be used to computed spinning gauged scalar and Dirac stars.  Section~\ref{sec3} discusses the boundary conditions and the  numerical method to obtain the solutions; then, the numerical results are presented. Besides the discussion on $g$, we find, for instance, that in both cases, the proportionality relation found in \cite{Herdeiro:2019mbz} between the angular momentum and the Noether charge (particle number) still holds.
Moreover, the ratio between the electric charge  and the Noether charge
is equal to the gauge coupling constant. In fact, some basic features are similar to those of the ungauged  stars.
In particular, for a given value of the gauge coupling constant $q$, one finds again a
limited range for the allowed frequencies of the matter fields,
which is bounded from above by
the fields' mass and from below by a  minimal frequency whose value
decreases with $q$.
Finally, in Section~\ref{sec4} we provide a discussion of the results and some final remarks.

\section{The general framework}
\label{sec2}

\subsection{The models and field equations}

We consider Einstein's gravity in 3+1 dimensions
minimally coupled with a spin-$s$ field
($s=0,1/2$) :
\begin{eqnarray}
\label{action}
\mathcal{S}=
\frac{1}{4\pi}
\int d^4 x \sqrt{-g}
\left [
\frac{R}{4G}
-\frac{1}{4}F^2
+
\mathcal{L}_{(s)}
\right] \ .
\end{eqnarray}
The notation and conventions used here follow closely
those in~\cite{Herdeiro:2017fhv,Herdeiro:2019mbz,Herdeiro:2020jzx}:
  $G$ is the gravitational constant, $R$ is the Ricci scalar associated with the
spacetime metric $g_{\mu\nu}$,
 $F_{\mu\nu} =\partial_\mu A_\nu - \partial_\nu A_\mu$ is the $U(1)$ field strength tensor.
For the matter Lagrangian
$\mathcal{L}_{(s)}$
we consider  two cases:
\begin{eqnarray}
\label{LS}
&& \mathcal{L}_{(0)}= - \frac{1}{2}g^{\alpha \beta}
\left[
(D_\alpha  \Phi)^* D_\beta \Phi
+
(D_\beta \Phi)^* D_\alpha \Phi
\right]
- \mu^2   \Phi^* \Phi \ , ~~{\rm with}~~ D_\nu \Phi=(\partial_\nu   + iq A_\nu) \Phi,
\\
%
\label{LD}
&&
 \mathcal{L}_{(1/2)} =-i
\left[
\frac{1}{2}
  \left( \{ \hat{\slashed D}  \overline{\Psi}   \} \Psi  -
     \overline{\Psi}  \hat{\slashed D}  \Psi
    \right)
+\mu \overline{\Psi}   \Psi
\right]\ ,~~{\rm with}~~\hat{D}_{\nu}  \Psi =(\partial_\nu - \Gamma_\nu + i q A_\nu)  \Psi ~.
\end{eqnarray}
Here, $\Phi$ is a complex scalar field;
$\Psi$ is a Dirac spinor, with four complex components.
For the scalar  field, the asterisk denotes complex conjugation;
$\overline{\Psi}$ denotes the Dirac conjugate \cite{Dolan:2015eua}.
 $\hat{\slashed D}\equiv \gamma^\mu \hat{D}_\mu$, where  $\gamma^\mu$ are the curved spacetime gamma matrices,
 and $\Gamma_\mu$
are the spinor connection matrices~\cite{Dolan:2015eua}.
In both cases, $\mu>0$ corresponds to the mass of the field(s),
while
$q$ is the gauge coupling constant.

\medskip

Variation of (\ref{action}) with respect to the metric leads to the Einstein field equations
\begin{eqnarray}
\label{Einstein-eqs}
G_{\alpha \beta} =2 G ~T_{\alpha \beta} \qquad {\rm with} \qquad T_{\alpha \beta}=  T_{\alpha \beta}^{(M)} +T_{\alpha \beta}^{(s)} \  ,
\end{eqnarray}
 where
$G_{\alpha \beta}$ is
the Einstein tensor and the pieces of the stress-energy tensor are
\begin{eqnarray}
\label{TM}
 T_{\alpha \beta}^{(M)}=F_{\alpha \gamma}F_{\beta \delta}g^{\gamma \delta}-\frac{1}{4}g_{\alpha\beta }F^2 \ ,
\end{eqnarray}
 for the Maxwell field,
 and the following
$T_{\alpha \beta}^{(s)}$  for the (gauged) scalar and Dirac fields, respectively:
\begin{eqnarray}
\label{TS}
&&
T_{\alpha \beta}^{(0)}=
(D_{ \alpha} \Phi)^* D_{ \beta}\Phi
+(D_{ \beta} \Phi)^* D_{ \alpha}\Phi
-g_{\alpha \beta}
                     \left[ \frac{1}{2} g^{\gamma \delta}
 (
(D_{ \gamma} \Phi)^* D_{ \delta}\Phi
    +
(D_{ \delta} \Phi)^* D_{ \gamma}\Phi
)+\mu^2   \Phi^* \Phi
                        \right] \ ,
%
%
\\
&&
T_{\alpha \beta}^{(1/2)}= -\frac{i}{2}
\left[
    \overline{\Psi}  \gamma_{(\alpha} \hat{D}_{\beta)} \Psi
-  \left\{ \hat{D}_{(\alpha} \overline{\Psi} \right\} \, \gamma_{\beta)} \Psi
\right]  \ .
\label{TD}
\end{eqnarray}
The corresponding matter field equations are:
\begin{eqnarray}
&&D_{\nu}D^{\nu}\Phi -\mu^2\Phi=0~~(s=0) \ , \qquad
(\gamma^\nu \hat{D}_\nu-\mu) \Psi =0  ~~(s=1/2) \ ,~~
\label{LP2}
\end{eqnarray}
and
 \begin{eqnarray}
\nabla_\alpha\mathcal{F}^{\alpha\beta}  =q J^\beta\ ,~~
{\rm with }~~
J^\beta=iq \big [ (D^{\beta}\Phi^*) \Phi-\Phi^*(D^\beta \Phi) \big ]~~~(s=0) \ ,  \quad {\rm or} \quad
~J^\beta= \overline{\Psi} \gamma^\beta \Psi  ~~~(s=1/2).
\label{Maxwelleq}
 \end{eqnarray}
for the Maxwell field.

These models are invariant under the $local$ $U(1)$ gauge transformation
\begin{eqnarray}
\label{gauge-transf}
(\Phi \to \Phi e^{-i q \alpha},~~\Psi \to \Psi e^{-i q \alpha}),~~{\rm and}~~A_\mu\to A_\mu +\partial_\mu \alpha,
\end{eqnarray}
with $\alpha$ a real function of spacetime coordinates.
The  current and the \textit{total} energy-momentum tensor are covariantly conserved,
\begin{eqnarray}
\nabla_\mu J^{\mu }= 0 \ , \qquad \nabla_\mu T^{\mu \nu} =0 \ .
\end{eqnarray}
 Then, integrating the timelike component of the 4-current $J^{\mu }$
on a spacelike hypersurface $\Sigma$
yields a conserved  \textit{Noether charge} (particle number):
\begin{eqnarray}
\label{Q}
Q =\int_{\Sigma}~J^t \ .
\end{eqnarray}

\subsection{The Ansatz}
\label{sec22}
The employed metric Ansatz is similar to that used in the  ungauged case \cite{Herdeiro:2019mbz},
with a line-element  possessing  two Killing vectors
 $ \partial_\varphi$
and
 $ \partial_t$
(with $\varphi$ and
$t$ the azimuthal and time
coordinate,
respectively):
\begin{eqnarray}
\label{metric}
ds^2=-e^{2F_0} dt^2+e^{2F_1}\left(dr^2+r^2 d\theta^2\right)+e^{2F_2}r^2 \sin^2\theta \left(d\varphi-\frac{W}{r} dt\right)^2\ .
\end{eqnarray}
This metric contains  four functions $(F_i;W)$, $i=0,1,2$,
which depend on the spherical coordinates $r$ and $\theta$ only.
The Minkowski spacetime background is approached for $r\to \infty$, where the asymptotic values are $F_i=0$, $W=0$.

In the scalar case, one the following Ansatz in terms of a single real function $\phi(r,\theta)$
and a complex phase ($\varphi,t$), with
\begin{eqnarray}
\label{S}
&&
\Phi=e^{i(m \varphi-w t)}\phi(r,\theta) \ .
\end{eqnarray}
In the case of a Dirac field, the Ansatz
 contains two complex functions \cite{Herdeiro:2019mbz}:
\begin{eqnarray}
\label{D}
\Psi = e^{i(m\varphi-w t) } \begin{pmatrix}
\psi_1 (r,\theta)
\\
\psi_2 (r,\theta)
\\
-i \psi_1^* (r,\theta)
\\
-i\psi_2^* (r,\theta)
\end{pmatrix}
\ ,~~
{\rm with}~~\psi_1 (r,\theta)=P(r,\theta)+i \mathcal{Q}(r,\theta)\ ,~~
\psi_2 (r,\theta)=X(r,\theta)+i Y(r,\theta)\ .
\end{eqnarray}
We shall employ the following orthonormal tetrad, as implied from the metric form (\ref{metric})
\begin{equation}
{\bf e}^0_\mu dx^\mu=e^{F_0}dt\ , \qquad {\bf e}^1_\mu dx^\mu=e^{F_1}dr\ , \qquad {\bf e}^2_\mu dx^\mu=e^{F_1}r d\theta\ ,
\qquad {\bf e}^3_\mu dx^\mu=e^{F_2}r \sin \theta \left(d\varphi-\frac{W}{r}dt\right) \ ,
\end{equation}
such that $ds^2=\eta_{ab}({\bf e}^a_\mu dx^\mu) ({\bf e}^b_\nu dx^\nu)$, where $\eta_{ab}={\rm diag}(-1,+1,+1,+1)$.

For a scalar field, the parameter $m$ in an integer, while for the Dirac field
  $m$ is a half-integer;
    $w$ is the field's frequency in both cases, which we shall take to be positive.
A study of each of the matter field equations in the far field reveals that
the solutions satisfy the bound state
condition $w<\mu$.

 The Ansatz for the $U(1)$ potential contains two real functions --
an electric and a magnetic potential, with:
\begin{eqnarray}
\label{ansatz-U1}
A= A_\mu dx^\mu= V(r,\theta) dt+ A_{\varphi}(r,\theta) r \sin \theta d\varphi \ .
\end{eqnarray}

Note that, in contrast to the ungauged case,  the $(t, \varphi)$-dependence of the scalar field $\psi$
can now be gauged away by applying the local $U(1)$ symmetry
(\ref{gauge-transf})
with $\alpha =  (m\varphi -\omega t)/q$.
However, this would also change the gauge field as $V\to V-\omega/q$,
$A_\varphi \to A_\varphi+m/q$, so that it would (formally) become singular in the $q\to 0 $ limit.
Therefore, in order
to be able to consider this limit, we prefer to keep the $(t,\varphi)$-dependence in
the  Ansatz and to fix
the corresponding gauge freedom by setting $V = A_\varphi = 0$ at infinity.

\subsection{Quantities of interest}

Given the above general Ansatz,
the computation of the explicit form of
the field equations
is straightforward.
Although the resulting  expressions are
in general
 too complicated to
include here,
the angular momentum density is simple enough, with
\begin{eqnarray}
\label{J0}
&&(T^{(0)})_\varphi^{t}=
2e^{-2F_0}
(m+q A_\varphi r \sin \theta)
 \left(w- q V-(m+q A_\varphi r \sin \theta) \frac{W}{r}\right)\phi^2\ ,
%
%
\\
&&
(T^{(1/2)})_\varphi^{t}=
e^{-F_0}(m+q A_\varphi r \sin \theta) (P^2+\mathcal{Q}^2+X^2+Y^2)
+
e^{-F_0-F_1+F_2}\sin \theta
\left\{
(PX+\mathcal{Q}Y)[1+r(F_{2,r}-F_{0,r})]  \right.
\\
&&{~~~~~} \left.
-\frac{1}{2}
(P^2+\mathcal{Q}^2-X^2-Y^2)
(\cot \theta+F_{2,\theta}-F_{0,\theta})
+2e^{-F_0+F_1} r \left(w-qV-(m+q A_\varphi r \sin \theta)\frac{ W}{r}\right)(\mathcal{Q}X-PY)
\right\} \ ,
\nonumber
\\
\label{JM}
&&
 (T^{(M)})_\varphi^{t}=
-\frac{e^{-2(F_0+F_1)}}{r}
\bigg[
\sin \theta  (r^2 A_{\varphi,r}V_{,r}+A_{\varphi,\theta}V_{,\theta})
+\sin^2 \theta W  (r^2 A_{\varphi,r}^2+A_{\varphi,\theta}^2)
\\
\nonumber
&&
{~~~~~~~~~~}
+A_\varphi
              \bigg(r \sin \theta  V_{,r}+\cos \theta V_{,\theta}
                            + W
\left (A_\varphi+2\sin \theta (r \sin \theta A_{\varphi,r}+\cos \theta A_{\varphi,\theta}
              ) \right) \bigg)
\bigg] \ .
\end{eqnarray}
Observe the presence of a $U(1)$-contribution in  $(T^{(s)})_\varphi^{t}$.
Of interest is also
 the temporal component of the current density:
\begin{eqnarray}
\label{Q0}
&&
J^{t}_{(0)}=2e^{-2F_0} \left(w - q V-(m+q A_\varphi r \sin \theta) \frac{ W}{r}\right) \phi^2 \ ,
\\
\label{Q12}
&&
J^{t}_{(1/2)}=2e^{-F_0}(P^2+\mathcal{Q}^2+X^2+Y^2)\ .
 \end{eqnarray}

 The ADM mass $M$ and the angular momentum $J$ of the solutions are read off from the asymptotic expansion:
\begin{eqnarray}
\label{asym}
g_{tt} =-1+\frac{2M}{r}+\dots\ , \qquad g_{\varphi t}=-\frac{2J}{r}\sin^2\theta+\dots \ . \ \ \
\end{eqnarray}
Of interest  is also the asymptotic decay of the gauge field
 \begin{eqnarray}
 \label{asym-matter-fields}
V\sim
\frac{Q_e}{r}+\dots~,~~~\qquad A_{\varphi}\sim \frac{\mu_m \sin \theta}{r^2}+\dots \ ,
 \end{eqnarray}
where
$Q_e$ and $\mu_m$ are the electric charge and the magnetic dipole moment, respectively.

The total angular momentum
 can also be computed as the integral of the corresponding density\footnote{The ADM mass can also be computed as volume integral; however, this is less relevant in the context of this work.}
\begin{equation}
\label{Js}
J = 2\pi \int^{\infty}_0 dr\, \int^{\infty}_0 d\theta r^2 e^{F_0+2F_1+F_2 }
\left(
(T^{(s)})_\varphi^{t}
+
(T^{(M)})_\varphi^{t}
\right)\ .
\end{equation}
The explicit form of the Noether charge,
as computed from~\eqref{Q},  is
\begin{equation}
\label{Qs}
Q \equiv  Q_{(s)}= 2\pi \int^{\infty}_0 dr\, \int^{\infty}_0 d\theta r^2 e^{F_0+2F_1+F_2 }j^{t}_{(s)} \ .
\end{equation}
For both a scalar field
and a Dirac one,
a straightforward computation
 shows that $J$, $Q$ and $Q_e$
are proportional,
\begin{equation}
\label{JQ}
J=m Q =\frac{m Q_e }{q} \  .
\end{equation}
Note that the above relation
is nontrivial,
since
 the angular momentum \textit{density} and Noether charge \textit{density} are $not$ proportional.
Nonetheless,
the proportionality still holds at the level of the integrated quantities.

As with any spinning system with gauge fields,
the solutions possess also a non-zero
  gyromagnetic ratio $g$,
    which defines how the magnetic dipole moment
is induced by the total angular momentum and charge, for a given total mass:
 \begin{equation}
 \mu_m =g\frac{Q_e}{2M}J \ .
 \label{gyro}
 \end{equation}
 
\section{The solutions}
\label{sec3}

\subsection{The boundary conditions and numerical method}

The numerical treatment of the problem is simplified by using
 some symmetries of the equations of motion \cite{Herdeiro:2017fhv}.
 Firstly, the factor of $ G$ in the Einstein field equations is set to unity by a redefinition of
 the matter functions,
\begin{equation}
\{\Phi,\Psi, A \}  \to \frac{1}{\sqrt{ G}}  \{ \Phi, \Psi , A \} \ .
\label{t-D}
\end{equation}
Secondly, one sets $\mu=1$
in the equations.
This can be done
  without any
loss of generality,
by noticing that
 the field equations remain invariant under the transformation
\begin{equation}
\label{sca1}
r \to \lambda r; ~~~~W \to \lambda W, ~ F_i \to F_i;~~~\{w,\mu, q\} \to \frac{1}{\lambda}\{w,\mu, q\}\ ;
 \ \ \
\left\{
\begin{array}{l}
\displaystyle{\Phi \to \Phi }
\\
\Psi \to  \frac{1}{\sqrt{\lambda}} \Psi
\\
A \to A
\end{array}
\right \},
\end{equation}
where $\lambda$ is a positive constant.
As for some quantities of interest, they transform as
\begin{eqnarray}
\label{sca2}
M\to \lambda M,~~
J\to \lambda^2 J,~~
Q_e\to \lambda Q_e,~~
Q\to \lambda^2 Q~~
{\rm and }~~
\mu_m\to \lambda \mu_m \ .
\end{eqnarray}
This  invariance is used to do the numerical work
 in units set by the field's mass,
$i.e.$ one takes $\lambda= {1}/{\mu}$.
Let us remark that only quantities which are invariant under the transformation
(\ref{sca1})
    (like
    $w/\mu$,
    $q/\mu$,
        $J/M^2$
    or $M \mu$)
    are relevant.

Given the  Ansatz
(\ref{S}),
(\ref{D}),
(\ref{ansatz-U1}),
all components of the energy momentum tensor are  zero,
except for
$
T_{rr},~T_{r\theta},~T_{\varphi \varphi},~T_{t t}
$
and
$T_{\varphi t}$,
which possess a $(r,\theta)$-dependence only
(although the scalar and spinor fields are $not$ time independent).

Then,
the Einstein field equations
with the energy momentum-tensors
\eqref{TM},
\eqref{TS},
\eqref{TD},
plus the matter field equations
 \eqref{LP2},
(\ref{Maxwelleq})
together with the Ansatz
(\ref{metric})
(\ref{S}),
(\ref{D}),
(\ref{ansatz-U1}),
lead to a system of seven (ten) coupled partial differential equations for the gauged  scalar (Dirac) models.
There are four equations for the metric functions $F_i,W$;
together with three (six) equations for the matter functions.
Apart from these, there are two constraint Einstein equations
which are not solved in practice,
being used the monitor
the accuracy of the numerical results.

The boundary conditions are found by considering an approximate construction of the solutions
on the boundary of the domain of integration
together with the assumption of regularity and asymptotic flatness.
The metric functions are subject to the following boundary conditions:
\begin{equation}
\partial_r F_i\big|_{r=0}=W\big|_{r=0}=0 \ , \qquad
 F_i\big|_{r=\infty}=W\big|_{r=\infty}=0 \ , \qquad 
\partial_\theta F_i\big|_{\theta=0,\pi}=\partial_\theta W\big|_{\theta=0,\pi}=0 \ .
 \end{equation}
The scalar field amplitude vanishes
on the boundary of the domain of integration,
\begin{equation}
\phi \big|_{r=0}= \phi \big|_{r=\infty}=\phi \big|_{\theta=0,\pi}=0 \ .
\end{equation}
 For a Dirac field, all solutions considered so far have
$m=1/2$, and satisfy the following boundary conditions
\begin{eqnarray}
&&
P\big |_{r=0}=   \mathcal{Q} \big |_{r=0}= X\big |_{r=0}=  Y\big |_{r=0}=0 \ , \qquad
 P\big |_{r=\infty}=\mathcal{Q} \big |_{r=\infty}=X\big |_{r=\infty}=Y\big |_{r=\infty}=0 \ ,
\\
\nonumber
&&
\partial_\theta P\big |_{\theta=0 }= \partial_\theta \mathcal{Q}\big |_{\theta=0 }=X\big |_{\theta=0 }=Y\big |_{\theta=0 }=0\ , \qquad
 P\big |_{\theta=\pi }=   \mathcal{Q}\big |_{\theta= \pi}=\partial_\theta X\big |_{\theta=\pi }= \partial_\theta Y\big |_{\theta=\pi }=0 \ .
\end{eqnarray}
Finally, for both $s=0,1/2$,
the Maxwell potentials satisfy the boundary conditions:
\begin{equation}
\partial_r V \big|_{r=0}=A_\varphi \big|_{r=0}=0\ , \qquad  V\big|_{r=\infty}=A_\varphi \big|_{r=\infty}=0 \ , \qquad
\partial_\theta  V \big |_{\theta=0,\pi }= \partial_\theta  A_\varphi \big |_{\theta=0,\pi }= 0 \ .
\end{equation}

After setting
$\mu=1$,
the problem has still three input parameters:
$\{w,m;q \}$ -- the field frequency, the azimuthal number and the gauge coupling constant.
The reported results in this work have $m=1$
for the scalar field and $m=1/2$ for a Dirac one.

The solutions are found by
using a fourth order
finite difference scheme.
The system of seven/ten equations is discretised on a grid with
$N_r\times N_\theta$ points; typically $N_r\sim 200$, $N_\theta \sim 50$.
We introduce
a new radial coordinate $x={r}/({r+c})$, which maps the semi-infinite region $[0,\infty )$ onto the unit interval
$[0,1]$, where $c$ is a constant of order one.

The gauged boson stars  were constructed by
using the professional package FIDISOL/CADSOL
 \cite{schoen}
 which uses a Newton-Raphson method.
The Einstein-Dirac-Maxwell
equations is solved with the Intel MKL PARDISO sparse direct solver \cite{pardiso},
and using the CESDSOL library. 
In all cases, the typical
errors are of order of $10^{-4}$.

Finally, we remark that the solutions shown here are fundamental states, with
 all matter functions being nodeless.
However, we predict the existence of
 a discrete set of solutions, indexed by the number of nodes, $n$,
of (some of) the matter function(s).

 {\small \hspace*{3.cm}{\it  } }
\begin{figure}[t!]
\hbox to\linewidth{\hss%
    \resizebox{9cm}{7cm}{\includegraphics{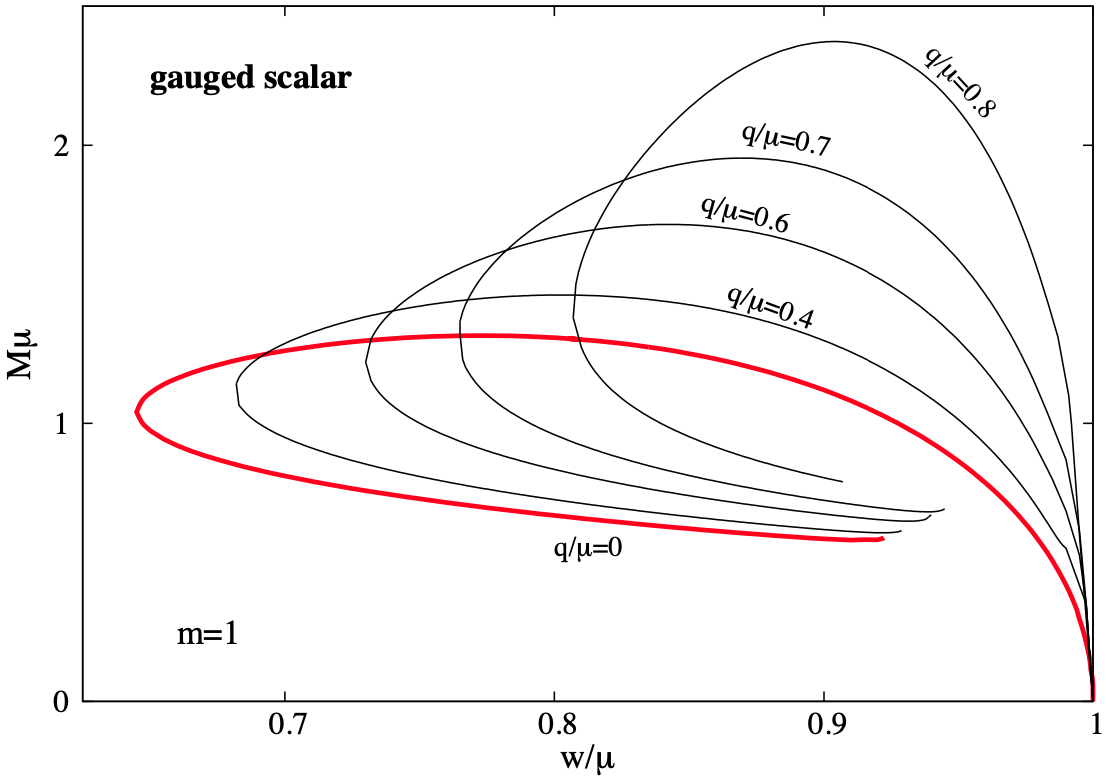}}
    \resizebox{9cm}{7cm}{\includegraphics{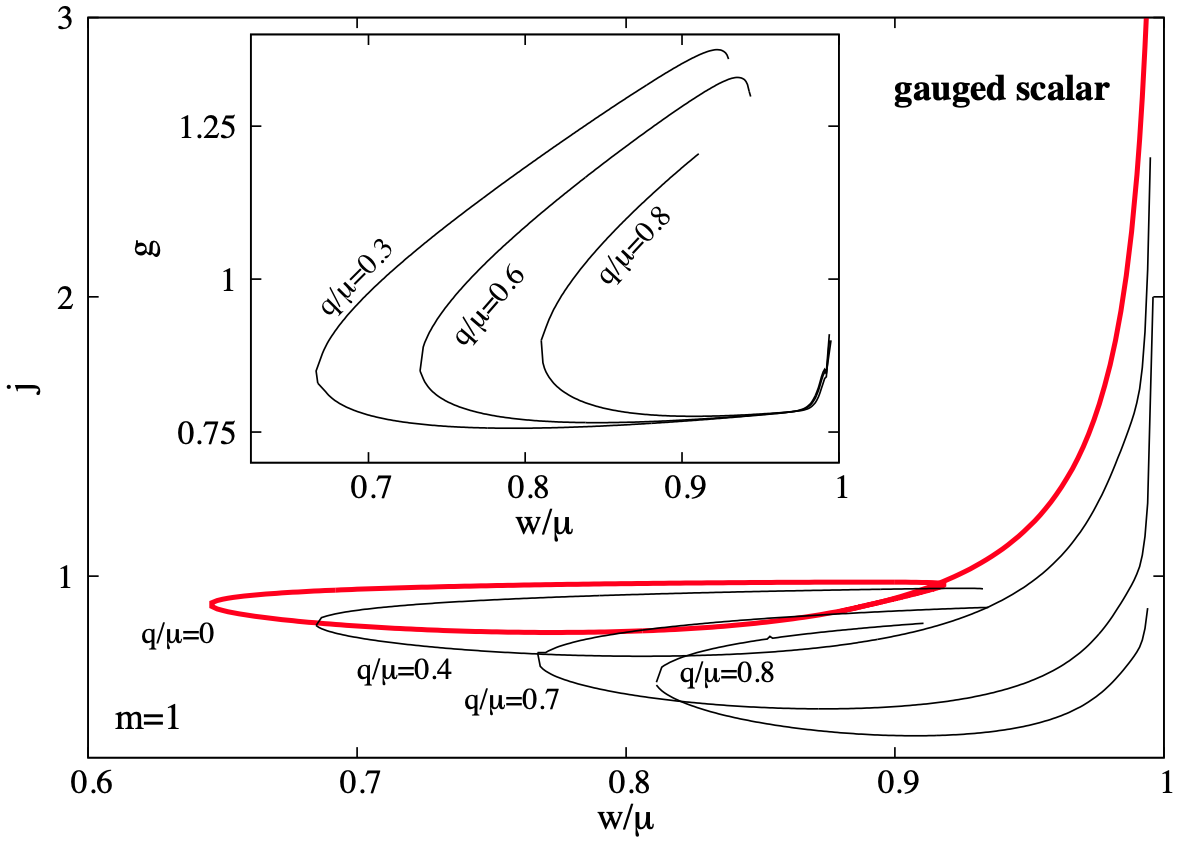}}
\hss}
\caption{\small
 The ADM mass $M$ (left panel) and the reduced angular momentum $j=J/M^2$
and the gyromagnetic ratio $g$ (right panel)
are shown for the families of
 spinning gauged  boson stars with
illustrative value of the gauged coupled constant $q$.
The quantities are given in units set by the field mass $\mu$.
}
\label{fig1}
\end{figure}

 {\small \hspace*{3.cm}{\it  } }
\begin{figure}[t!]
\hbox to\linewidth{\hss%
    \resizebox{9cm}{7cm}{\includegraphics{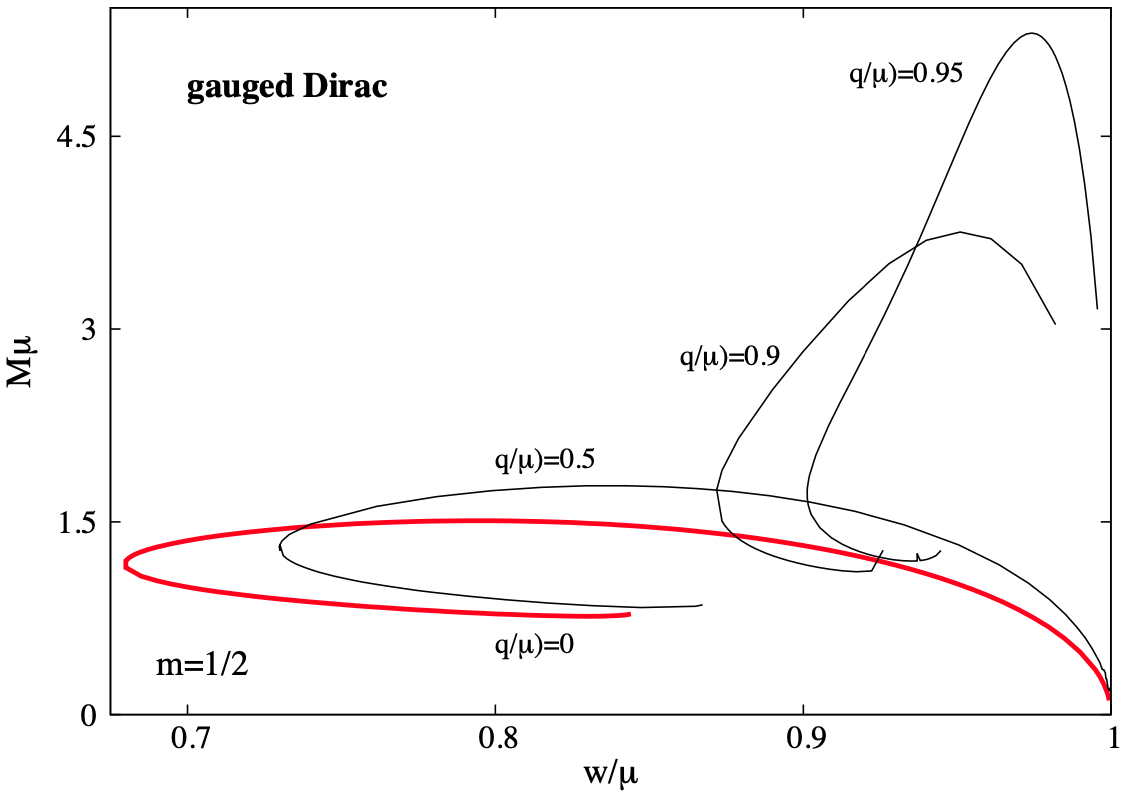}}
    \resizebox{9cm}{7cm}{\includegraphics{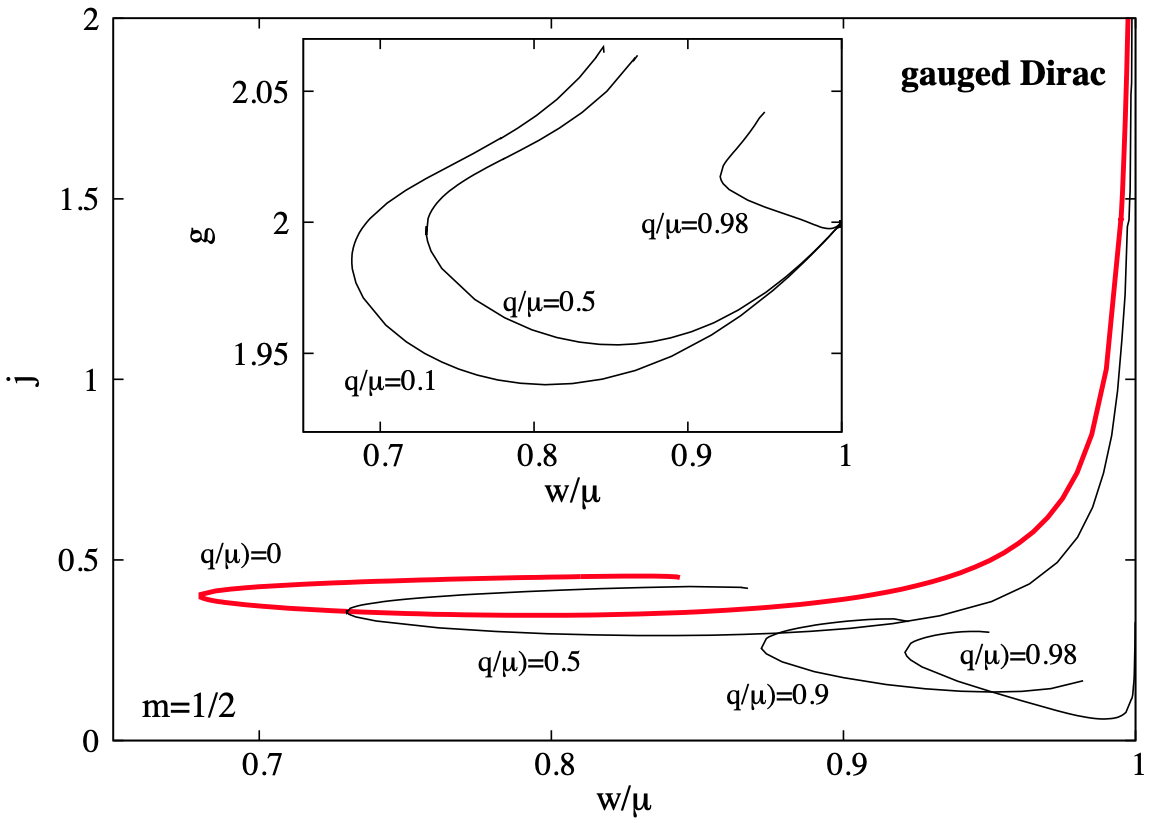}}
\hss}
\caption{\small
Same as Figure \ref{fig1}
for spinning  gauged Dirac stars.
}
\label{fig2}
\end{figure}

\subsection{Numerical results  }

In our approach, we start with the ungauged solution in~\cite{Herdeiro:2019mbz}
($i.e$ $q=0$ and $A_\varphi=V=0$).
Then one can smoothly turn on the gauge field by increasing (from zero)
the value of the gauge coupling constant
$q$, while keeping fixed the other input constants (in particular the parameters $w,m$).
The basic properties of the gravitating spinning gauged boson
and Dirac
stars solutions so constructed can be summarized as follows:

For a given values of $w$, spinning solutions
appear to
exist up to a maximal value of the gauge coupling constant
only, $q_{\rm max}$, where the numerical process stops to converge. We remark that all global charges stay finite in that limit.
The physical  mechanism behind this behaviour is
likely  similar to that discussed for the spherically
symmetric case
\cite{Jetzer:1989av,Finster:1998ux}:
the electric charge repulsion becomes too strong
and localized solutions cease to exist.
A precise determination of $q_{\rm max}$ is challenging;
all solutions found so far have
$q/\mu <1$.

Given a value of $q$, the full spectrum of solutions is
constructed
by varying the field frequency $w$.
The gauged spinning stars
exist for a limited range of frequencies
$0<w_{\rm min}<w<w_{\rm max}=\mu$,
see
 Figs \ref{fig1}, \ref{fig2}. Observe that
the minimal frequency increases with $q$.

A backbending towards
larger values of $w$  is observed as $w\to w_{min}$,
for both $s=0,1.2$.
One may expect that,
similar to the spherically symmetric case, this backbending would
lead to an inspiraling of the solutions towards a limiting configuration
with $w_c>w_{\rm min}$.
However, the construction of these secondary branched is a complicated
numerical task,
 which we do not attempt in this work.
Also, the numerical accuracy decreases as $w/\mu\to 1$,
with a delocalization of the profiles for the scalar  and spinor functions,
a different approach being necessary for the study of this limit.

As seen in
Figures  \ref{fig1}, \ref{fig2}.
for any $q$, the $(w,M)$
looks qualitatively similar to that found in the ungauged case ($q=0$). The observed trend is that
the maximal value of $M$ increases with $q$.
Note that
a similar behaviour is found for the  $(w,J)$-dependence.
Also, the minimal value of the
 reduced angular momentum $j=J/M^2$
decreases with $q$.

The shape of the metric functions and of the $s=0,1/2$ matter functions
is rather similar to the ungauged case.
Concerning the gauge field, the electric potential $V$
does not possess a
 strong angular dependence;
however, the magnetic potential $A_\varphi$
exhibits an involved angular dependence.

\begin{figure}[t!]
\hbox to\linewidth{\hss%
    \resizebox{6cm}{3cm}{\includegraphics{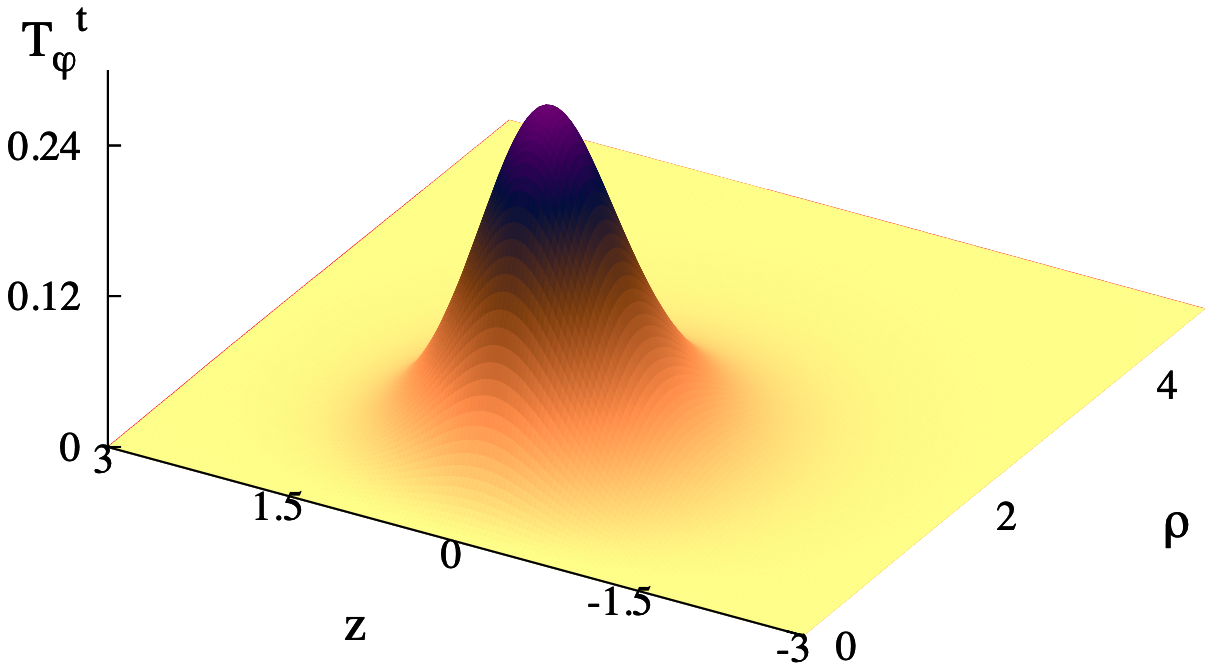}}
    \resizebox{6cm}{3cm}{\includegraphics{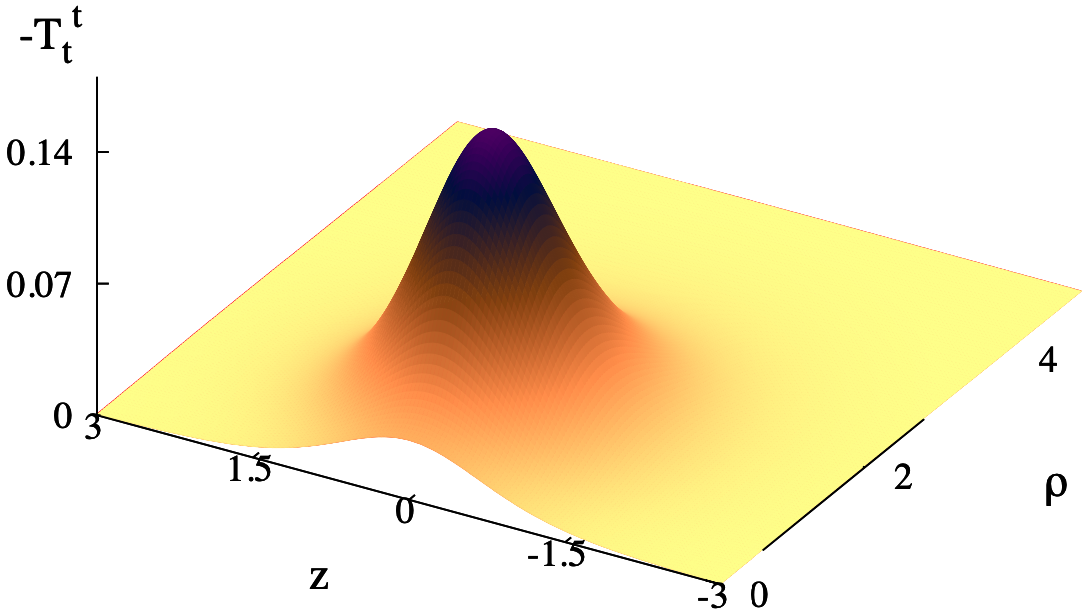}}
    \resizebox{6cm}{3cm}{\includegraphics{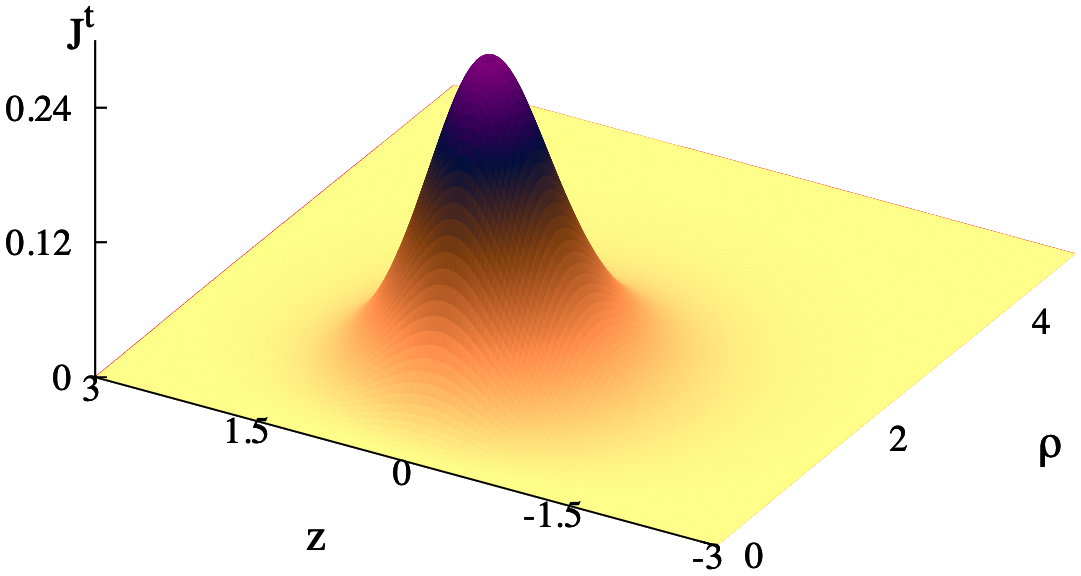}}
\hss}
\caption{\small
The components
$T_\varphi^t$
and
$T_t^t$
of the total energy momentum tensor, associated with angular momentum and energy densities 
and the $J^t$-component of the current are shown as a function
of the cylindrical coordinates $(\rho,z)$
(with $\rho=r\sin \theta$, $z=r \cos \theta$)
for a typical spinning gauged boson star.
The input parameters are $m=1$, $w =0.75$, $\mu=1$ and $q=0.5$.
}
\label{fig3Ds}
\end{figure}

\begin{figure}[t!]
\hbox to\linewidth{\hss%
    \resizebox{6cm}{3cm}{\includegraphics{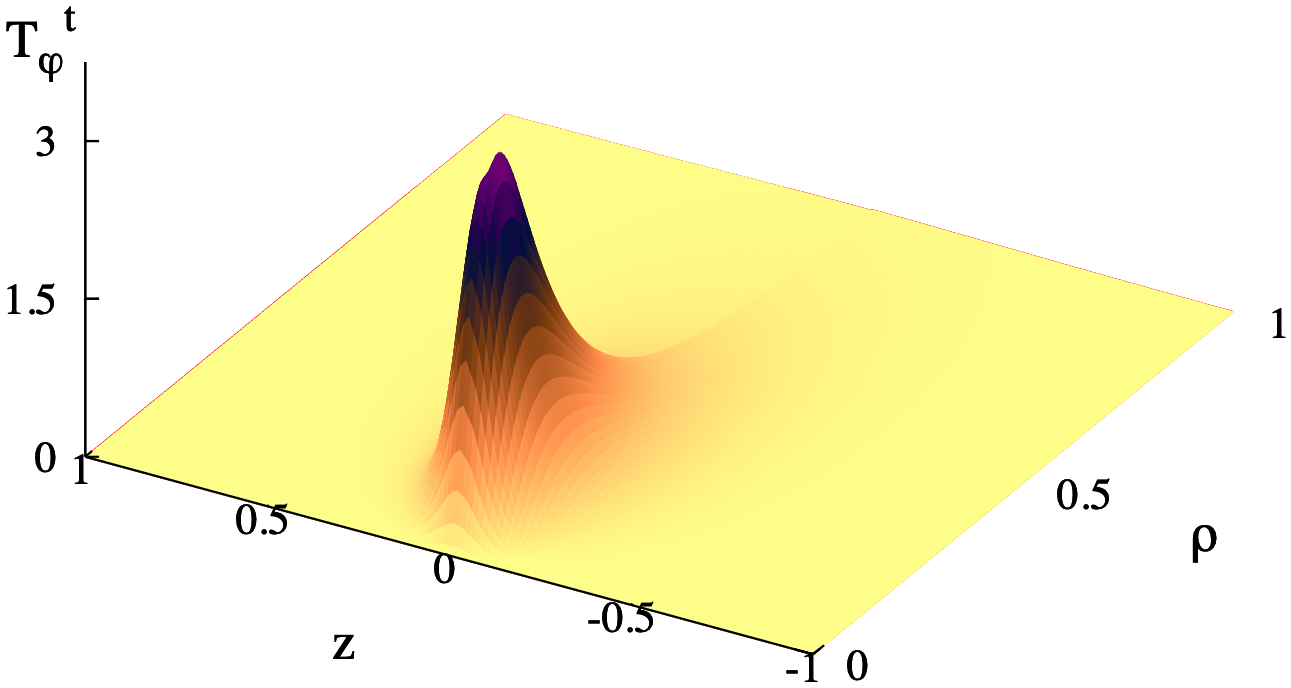}}
    \resizebox{6cm}{3cm}{\includegraphics{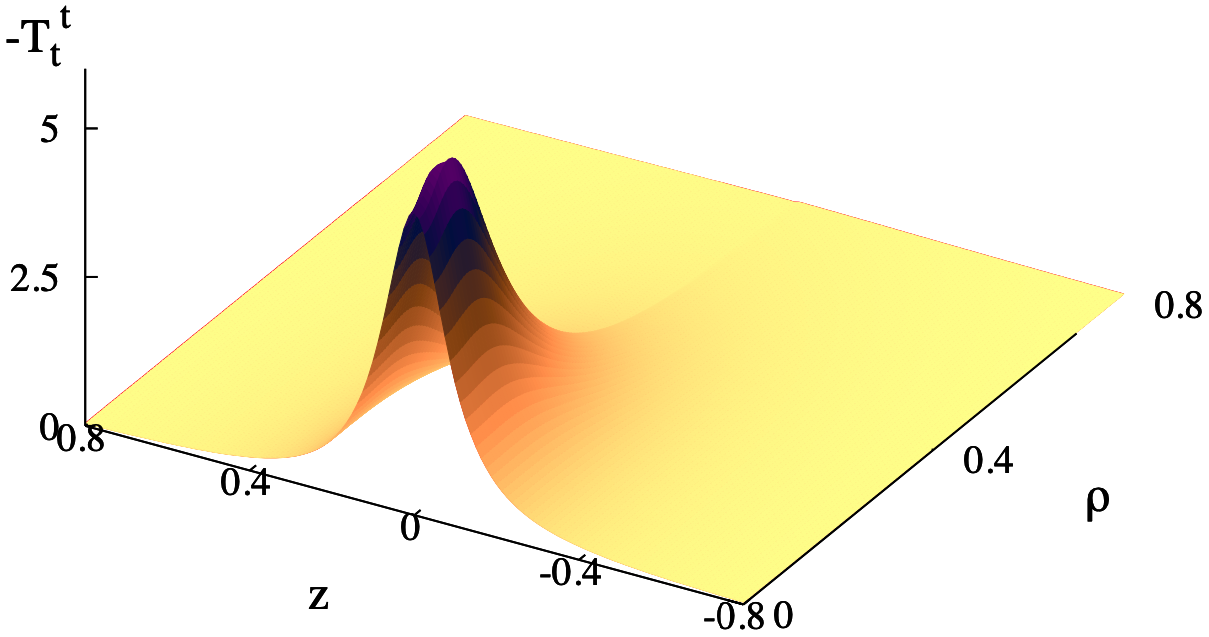}}
    \resizebox{6cm}{3cm}{\includegraphics{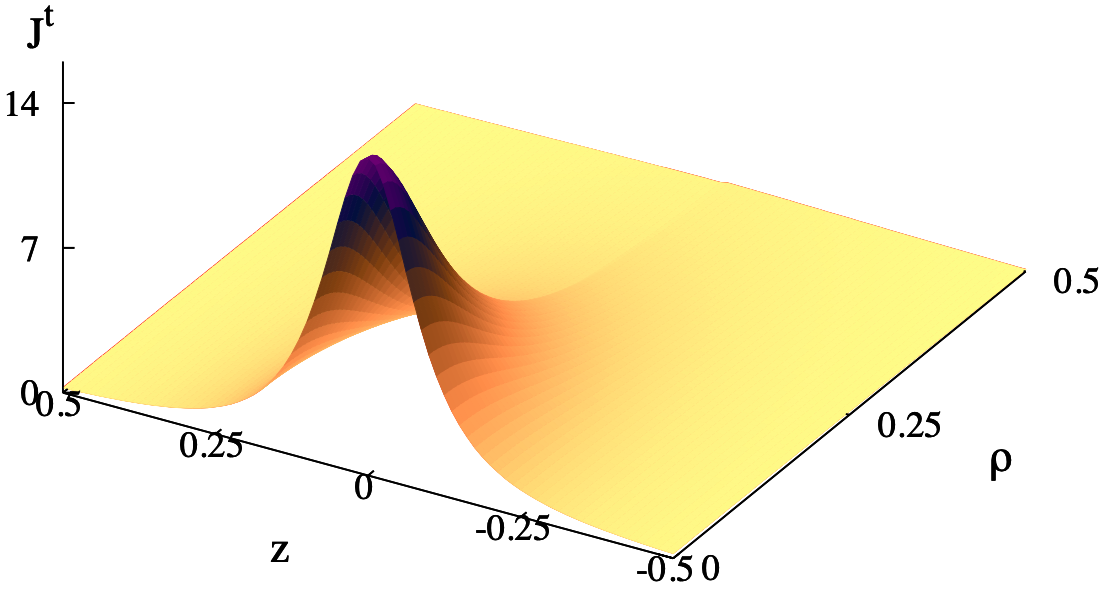}}
\hss}
\caption{\small
Same as Figure \ref{fig3Ds}
for a spinning  gauged Dirac stars
with $m=1/2$
and
the same values of $w,q$ and $\mu$.
}
\label{fig3Dd}
\end{figure}

Also, as seen in Figure {\ref{fig3Ds}}
the energy density $-T_t^t$ of the $s=0$ solutions
 is localized in a finite region in  the equatorial plane
and decreases monotonically along the symmetry axis,
such that the typical energy
density isosurfaces have a toroidal shape.
 At the same time, the angular momentum density
(which equals the Noether charge density)
has a strong peak in the equatorial plane.
Note  that while $T_\varphi^t$ vanishes on the symmetry axis,
this is not the case for $T_t^t$.
The energy  density  distribution is
still toroidal for typical Dirac stars,
although becoming more spheroidal than in the scalar case - see Figure {\ref{fig3Dd}}.

The gyromagnetic ratio
$g$
of the solutions has a nontrivial dependence on both
frequency and gauge coupling constant,
taking values around 1
(for $s=0$
stars)
and 2 for
(for $s=1/2$
stars), the larger deviations being found along the secondary branches of solutions
(see  the insets in Figures~\ref{fig1}, \ref{fig2}, right panels).
Interestingly, the extrapolation of the numerical results 
towards the Newtonian limit $w/\mu \rightarrow 1$
suggest that
$g\to 1$ 
$(s=0)$
and
$g\to 2$
$(s=1/2$),
independently of the value of the gauge coupling constant $q$.

\begin{figure}[h!]
\begin{center}
\includegraphics[width=0.465\textwidth]{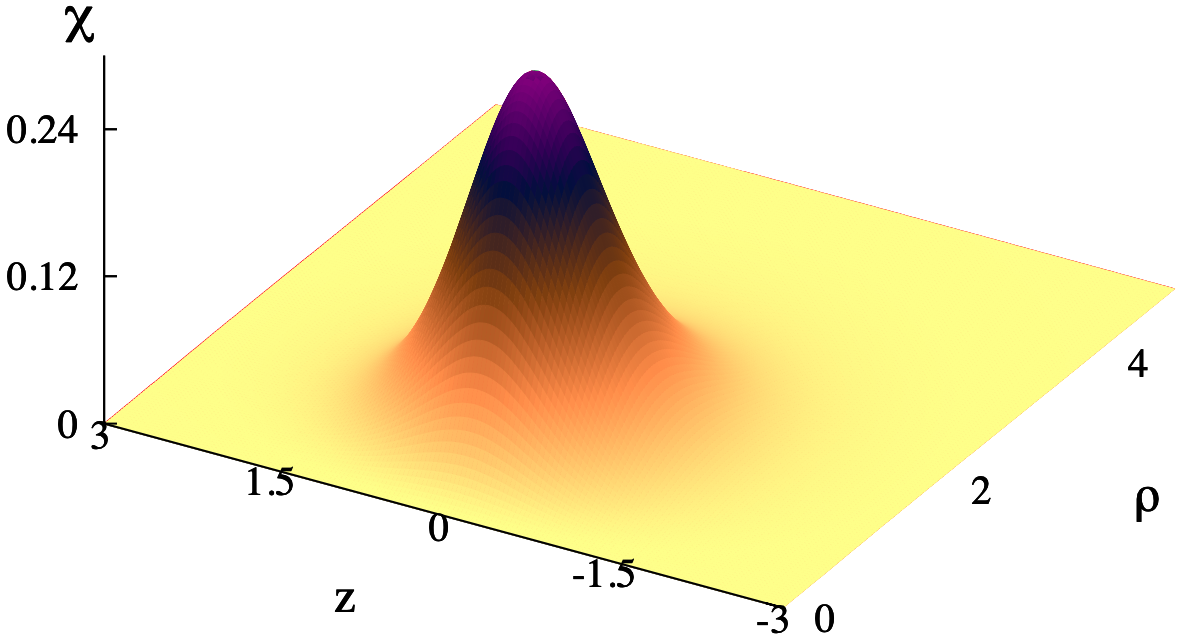}
\includegraphics[width=0.465\textwidth]{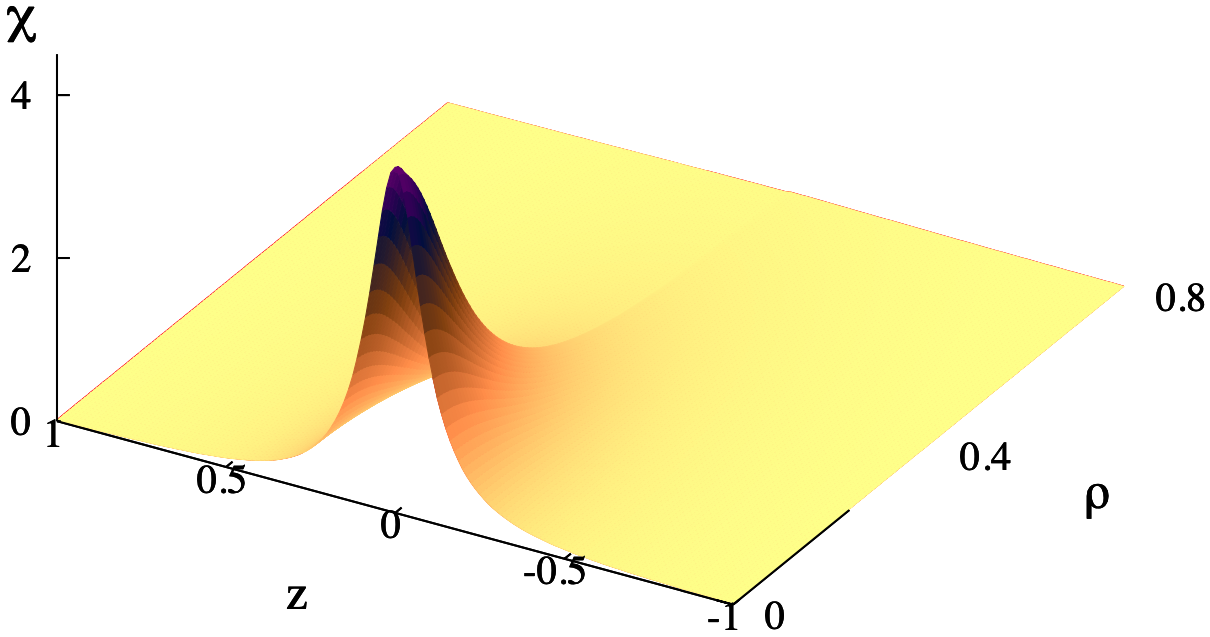}
\caption{\small{
The quantity $\chi= \left (T_{\mu\nu} - \frac12 T g_{\mu\nu}\right)X^\mu X^\nu  $
(with  the timelike vector $X^\mu$, $X^\mu X_\mu=-1$)
is shown for the
a  spinning gauged boson  (left panel) star
and a Dirac (right panel) star, which correspond to the solutions
  in Figures \ref{fig3Ds} and \ref{fig3Dd}.
The strong energy condition $\chi \geq 0$
is satisfied in both cases.
}}
\label{Ttot}
\end{center}
\end{figure}

It is also of interest to study the strong energy condition
\begin{eqnarray}
\chi= \left (T_{\mu\nu} - \frac12 T g_{\mu\nu}\right)X^\mu X^\nu \ge 0
\end{eqnarray}
(with  the timelike vector $X^\mu$, $X^\mu X_\mu=-1$).
We have monitored this condition for a number of solutions and
have found that $\chi>0$ in all cases
(see  Figure (\ref{Ttot}) where this quantity is shown for the same
configurations as in Figures (\ref{fig3Ds}),  (\ref{fig3Dd})).

 {\small \hspace*{3.cm}{\it  } }

 {\small \hspace*{3.cm}{\it  } }

\begin{figure}[h!]
\begin{center}
\includegraphics[width=0.495\textwidth]{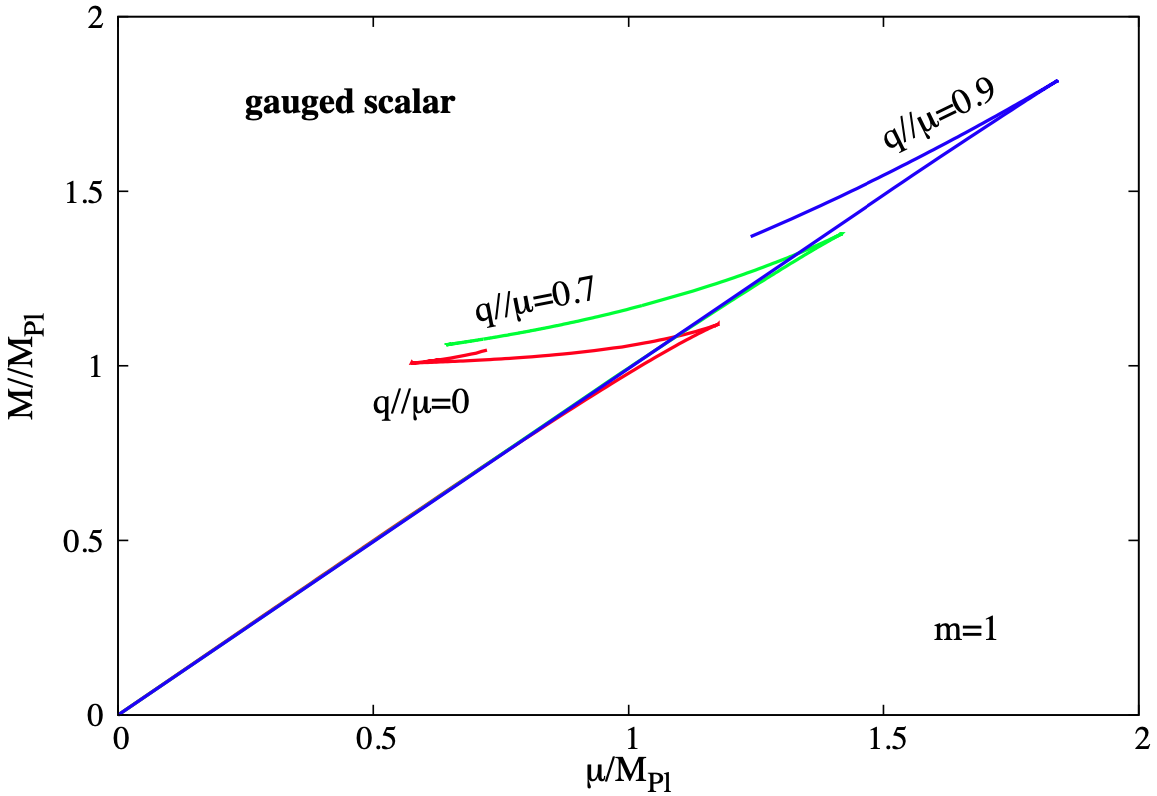}
\includegraphics[width=0.495\textwidth]{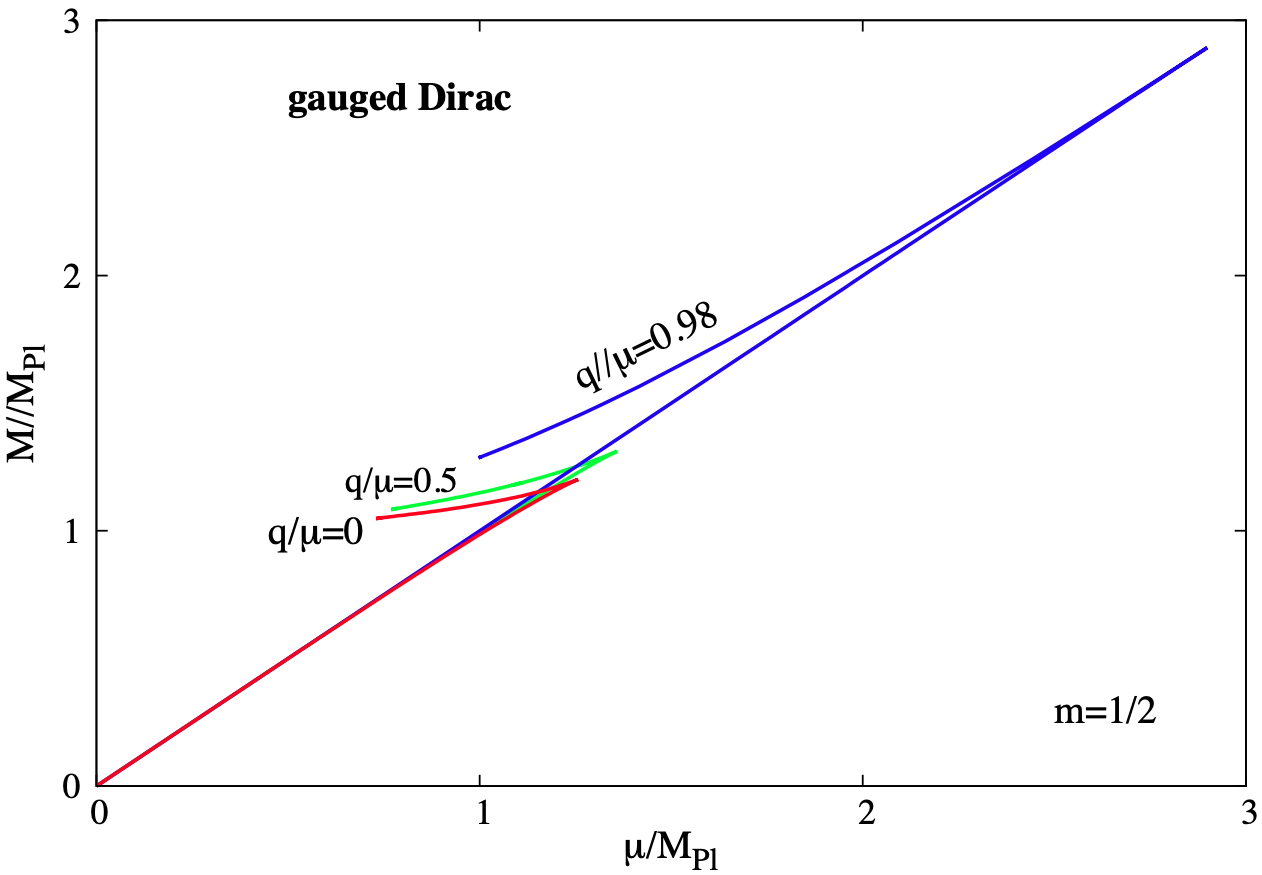}
\caption{\small{
(Left panel) The mass $M$ $vs.$  scalar field mass $\mu$,
in Planck units is shown for the three values of the invariant ratio $q/M$
(where $q$ is the gauge coupling constant).
(Right panel) Same for the gauged Dirac case.
In both cases, the single particle condition $Q=1$ is imposed.
}}
\label{fig4}
\end{center}
\end{figure}

\subsection{ The one particle picture}

The results above are found for a classical treatment of the fields. 
In particular, the particle number
is arbitrary and results from the numerical output
for some given physical parameters $(w,\mu,q)$.
If one tries to go beyond the classical field theory analysis and impose the quantum nature of fermions, this
requires $Q=1$ for Dirac stars.
 This condition can also be imposed for boson stars, although in this case it is not a mandatory requirement.

As noticed in the original work \cite{Finster:1998ws}, the one particle condition
can be imposed by making use of a scaling symmetry of the equations. 
That is, given a numerical solution with $Q^{\rm (num)}$,
one uses  (\ref{sca1})
with $\lambda=\sqrt{Q^{\rm (num)}}$, such that the scaled solution has $Q=1$ and
$Q_e=q$.
Then, as discussed in~\cite{Herdeiro:2017fhv},
the $(w,M)$-curves in Figures  \ref{fig1}, \ref{fig2}.
are not sequences of solutions with constant $\mu$, $q$ and varying $M$ (and $Q$); rather,
 it is a sequence with constant $Q=1$ and varying $\mu$ and $q$.
Thus, since $\mu$, $q$ are parameters in the action,
the curves would correspond to sequences of solutions of different models.

The resulting picture is shown in Figure \ref{fig4},
where we plot same  of the data as  in Figure~\ref{fig1}
but imposing the single particle condition.
One can see  that  the maximal mass for both the solutions' mass and field mass is
of order of the Planck mass.

\section{Conclusions}
\label{sec4}

The main purpose of this work was to provide a comparative analysis
of two different types of solitonic solutions of GR-matter systems,
with  matter fields of spin
$0$  and $1/2$, respectively. Here, and
different from the previous study in~\cite{Herdeiro:2019mbz},
the scalar and Dirac fields are gauged,
with a $local$
$U(1)$ symmetry.

We have confirmed that,
as classical field theory solutions,
gauging the fields still does not lead to
a clear
 distinction between the fermionic/bosonic nature of the field,
the
configurations, which still
possess a variety of  similar features. 
Interestingly, the gyromagnetic ratio of the solutions appears as a distinguishing feature, with values around 1 for the scalar stars and 2 for the Dirac stars.
It would be interesting to study also spinning gauged  Proca stars and in particular consider their gyromagnetic ratio

Finally, let us remark on another difference between the models. 
 The gauged spinning scalar boson stars can be in equilibrium 
with a black hole horizon~\cite{Delgado:2016jxq}. 
This still does not seem to be possible for the Dirac case.

\section*{Acknowledgements}

 The work of C.H. and E.R. has been supported  by the Center for Research and
 Development in Mathematics and Applications (CIDMA) through the Portuguese Foundation
 for Science and Technology (FCT - Funda\c c\~ao para a Ci\^encia e a Tecnologia),
 references UIDB/04106/2020 and UIDP/04106/2020 and by national funds (OE),
 through FCT, I.P., in the scope of the framework contract foreseen in the
 numbers 4, 5 and 6 of the article 23, of the Decree-Law 57/2016, of August 29,
 changed by Law 57/2017, of July 19. We acknowledge support from the projects
 PTDC/FIS-OUT/28407/2017,  CERN/FIS-PAR/0027/2019 and PTDC/FIS-AST/3041/2020.
 The authors would like to acknowledge networking support by the COST Action CA16104.
Y.S. gratefully acknowledges support by the Ministry of
Education of Russian Federation, project FEWF-2020-0003.
This work has further been supported by the European Union's Horizon 2020
research and innovation (RISE) programme H2020-MSCA-RISE-2017 Grant No.~FunFiCO-777740.

 \begin{small}
 
 \end{small}


\begin{thebibliography}{99}

\bibitem{Herdeiro:2017fhv}
  C.~A.~R.~Herdeiro, A.~M.~Pombo and E.~Radu,
  Phys.\ Lett.\ B {\bf 773} (2017) 654
  [arXiv:1708.05674 [gr-qc]].
%
\bibitem{Herdeiro:2019mbz}
C.~Herdeiro, I.~Perapechka, E.~Radu and Y.~Shnir,
Phys. Lett. B \textbf{797} (2019), 134845
[arXiv:1906.05386 [gr-qc]].

\bibitem{Herdeiro:2021lwl}
C.~A.~R.~Herdeiro, A.~M.~Pombo, E.~Radu, P.~Cunha, V.P. and N.~Sanchis-Gual,
JCAP \textbf{04} (2021), 051
[arXiv:2102.01703 [gr-qc]].

\bibitem{Bustillo:2020syj}
J.~C.~Bustillo, N.~Sanchis-Gual, A.~Torres-Forn\'e, J.~A.~Font, A.~Vajpeyi, R.~Smith, C.~Herdeiro, E.~Radu and S.~H.~W.~Leong,
Phys. Rev. Lett. \textbf{126} (2021) no.8, 081101
[arXiv:2009.05376 [gr-qc]].


\bibitem{Kaup:1968zz}
D.~J.~Kaup,
Phys. Rev. \textbf{172} (1968), 1331-1342

\bibitem{Ruffini:1969qy}
R.~Ruffini and S.~Bonazzola,
Phys. Rev. \textbf{187} (1969), 1767-1783


\bibitem{Jetzer:1989av}
  P.~Jetzer and J.~J.~van der Bij,
  Phys.\ Lett.\ B {\bf 227} (1989) 341.
\bibitem{Pugliese:2013gsa}
  D.~Pugliese, H.~Quevedo, J.~A.~Rueda H. and R.~Ruffini,
  Phys.\ Rev.\ D {\bf 88} (2013) 024053
  [arXiv:1305.4241 [astro-ph.HE]].



\bibitem{Schunck:1996wa}
  F.~E.~Schunck and E.~W.~Mielke,
  Phys.\ Lett.\ A {\bf 249} (1998) 389.
\bibitem{Yoshida:1997qf}
  S.~Yoshida and Y.~Eriguchi,
  Phys.\ Rev.\ D {\bf 56} (1997) 762.
\bibitem{Kleihaus:2005me}
  B.~Kleihaus, J.~Kunz and M.~List,
  Phys.\ Rev.\  D {\bf 72} (2005) 064002
  [arXiv:gr-qc/0505143].
\bibitem{Kleihaus:2007vk}
  B.~Kleihaus, J.~Kunz, M.~List and I.~Schaffer,
  Phys.\ Rev.\ D {\bf 77} (2008) 064025
  [arXiv:0712.3742 [gr-qc]].


\bibitem{Delgado:2016jxq}
J.~F.~M.~Delgado, C.~A.~R.~Herdeiro, E.~Radu and H.~Runarsson,
Phys. Lett. B \textbf{761} (2016), 234-241
[arXiv:1608.00631 [gr-qc]].

\bibitem{Brihaye:2009dx}
  Y.~Brihaye, T.~Caebergs and T.~Delsate,
  arXiv:0907.0913 [gr-qc].
\bibitem{Collodel:2019ohy}
L.~G.~Collodel, B.~Kleihaus and J.~Kunz,
Phys. Rev. D \textbf{99} (2019) no.10, 104076
[arXiv:1901.11522 [gr-qc]].



\bibitem{Kleihaus:2016rgf}
B.~Kleihaus, J.~Kunz and F.~Navarro-Lerida,
Class. Quant. Grav. \textbf{33} (2016) no.23, 234002
[arXiv:1609.07357 [hep-th]].

\bibitem{Finster:1998ws}
F.~Finster, J.~Smoller and S.~T.~Yau,
Phys. Rev. D \textbf{59} (1999), 104020
[arXiv:gr-qc/9801079 [gr-qc]].


\bibitem{Finster:1998ux}
F.~Finster, J.~Smoller and S.~T.~Yau,
Phys. Lett. A \textbf{259} (1999), 431-436
[arXiv:gr-qc/9802012 [gr-qc]].

\bibitem{Carter:1968rr}
B.~Carter,
Phys. Rev. \textbf{174} (1968), 1559-1571

\bibitem{Morel2021}
L.~Morel, Z-~Yao, P.~Clad\'e  et al.
Nature {\bf 588} (2020)  61
\bibitem{Garfinkle:1990ib}
D.~Garfinkle and J.~H.~Traschen,
Phys. Rev. D \textbf{42} (1990), 419-423
\bibitem{Khriplovich:1997ni}
I.~B.~Khriplovich and A.~A.~Pomeransky,
J. Exp. Theor. Phys. \textbf{86} (1998), 839-849
\bibitem{Pfister:2002dz}
H.~Pfister and M.~King,
Phys. Rev. D \textbf{65} (2002), 084033
\bibitem{Pfister:2002}H.~Pfister and M.~King,
  Class.\ Quant.\ Grav.\   {\bf 20} (2002), 205.




\bibitem{Herdeiro:2020jzx}
C.~A.~R.~Herdeiro and E.~Radu,
Symmetry \textbf{12} (2020) no.12, 2032
[arXiv:2012.03595 [gr-qc]].



\bibitem{Jetzer:1993nk}
  P.~Jetzer, P.~Liljenberg and B.~S.~Skagerstam,
  Astropart.\ Phys.\  {\bf 1} (1993) 429
  [astro-ph/9305014].


\bibitem{Dolan:2015eua}
  S.~R.~Dolan and D.~Dempsey,
  Class.\ Quant.\ Grav.\  {\bf 32} (2015) no.18,  184001
  [arXiv:1504.03190 [gr-qc]].

\bibitem{Brill:1957fx}
  D.~R.~Brill and J.~A.~Wheeler,
  Rev.\ Mod.\ Phys.\  {\bf 29} (1957) 465.

\bibitem{Dolan:2009kj}
  S.~Dolan and J.~Gair,
  Class.\ Quant.\ Grav.\  {\bf 26} (2009) 175020
  [arXiv:0905.2974 [gr-qc]].

\bibitem{Soler:1970xp}
  M.~Soler,
  Phys.\ Rev.\ D {\bf 1} (1970) 2766.
\bibitem{Finkelstein:1951zz}
  R.~Finkelstein, R.~LeLevier and M.~Ruderman,
  Phys.\ Rev.\  {\bf 83} (1951) 326.





\bibitem{schoen}
 W. Sch\"onauer and R. Wei\ss ,
 J. Comput. Appl. Math. 27, 279 (1989) 279;
 \\
 M. Schauder, R. Wei\ss\ and W. Sch\"onauer,
 Universit\"at Karlsruhe, Interner Bericht Nr. 46/92 (1992).
\bibitem{pardiso}
N.I.M.~Gould,  J.A.~Scott and Y.~Hu,
ACM Transactions on Mathematical Software {\bf 33}~(2007)~10;
\\
O.~Schenk and K.~G\"artner
Future Generation Computer Systems \textbf{20} (3) (2004) 475.



\end{thebibliography}
\end{document}